\documentclass[aps,preprint]{revtex4}
\usepackage{epsfig}
\usepackage{amsmath}
\usepackage{epsfig}

\begin{document}
	\title
	{Solvable models of an open well and  a bottomless barrier: 1-D potentials}
	\author{Zafar Ahmed$^1$, Dona Ghosh$^2$, Sachin Kumar$^3$, Nihar Turumella$^4$}
\affiliation{$^1$Nuclear Physics Division, Bhabha Atomic Research Centre, Mumbai 400 085, India\\
	$^2$Depatment of Mathematics, Jadavpur University, Kolkata 700032, India \\
	$^3$Theoretical Physics Section, Bhabha Atomic Research Centre, Mumbai 400 085, India\\
	$^4$Visvesvaraya National Institute of Technology, Nagpur 440010, India}
	\email{1:zahmed@barc.gov.in, 2:rimidonaghosh@gmail.com, 3:Sachinv.barc.gov.in, 4:nihar.turumell@gmail.com}
		  
	\date{\today}
	\begin{abstract}
		We present  one dimensional potentials $V(x)= V_0[e^{2|x|/a}-1]$ as  solvable models of a well $(V_0>0)$ and a barrier ($V_0<0$). Apart from being  new addition to solvable models, these models are instructive for finding bound and scattering states from the analytic solutions of Schr{\"o}dinger equation. The exact analytic (semi-classical and quantal) forms for bound states of the well and reflection/transmission $(R/T)$  coefficients for the barrier have been derived. Interestingly, the crossover energy $E_c$ where $R(E_c)=1/2=T(E_c)$  may occur below/above or at the barrier-top.
		A connection between  poles of these coefficients and bound state eigenvalues of the well has also been demonstrated.		
	\end{abstract}

\maketitle
\section{Introduction}
Solutions of Schr{\"o}dinger equation arising from dimensional potentials have been  enriching the understanding of microscopic world through quantum mechanics [1-5]. At the heart of Planck's explanation of black body radiation and the Einstein's theory  of  specific heat of solids there lies the harmonic oscillator potential $V(x)=\frac{1}{2}\mu\omega^2 x^2$ and its discrete bound state spectrum $(n+1/2)\hbar \omega$. Transmission across  a triangular potential barrier explains cold emission of electrons from metals through Fowler-Nordheim factor. Tunneling through a parabolic barrier could explain nuclear fusion cross-sections. Transmission at a simple rectangular barrier is Gamow's model of $\alpha$ decay from nucleus. In microscopic world, the success story
of one dimensional potential wells and barriers, which have one minimum and one   
maximum respectively is most interesting.

However, there are only few potential functions which are amenable to exact analytic solutions. Textbooks discuss potentials like Dirac-Delta [1], rectangular (square) [1-4], parabolic [4,5], triangular [5], Eckart [3,4], Fermi-step  [3,4] and Rosen-Morse [6] potentials. The exact solvability of Scarf II [7] and the versatile Ginocchio's potential [8] has come up  rather late. The 
Exponential potential $V(x)=-V_0 e^{x/a}$ has been suggested [10] as a simple practice problem that gives rise to the simplest forms for reflection $R(E)$ and transmission $T(E)$ coefficients.  The Morse [11] and one more potential barrier [12] have been presented giving rise to simple analytic expressions  for $T(E)$ and $R(E)$. Symmetric exponential [9], bi-harmonic [13] and symmetric triangular [5] potentials  have been solved in terms of  higher order functions, like Bessel, parabolic cylindrical and Airy functions. These models are even more welcome now, as various packages can calculate higher order functions fast and accurate. Though quick numerical algorithms of integration of Sch{\"o}dinger equation are also available yet handing higher order functions to extract bound and scattering states analytically or semi-analytically cannot loose its charm and importance specially when potentials diverge to $\pm \infty$ as $|x| \rightarrow \infty.$ This is so because for scattering states,  asymptotic boundary conditions cannot be put unless the
Schr{\"o}dinger equation is analytically solvable.

The alternate quantum mechanical approaches like super symmetric methods [14,15] have furthered the pursuit of exactly solvable models remarkably. Very interesting analytic forms [16,17] of complex scattering amplitudes $r(k)$ and $t(k)$ for Scarf II have emerged from such studies. Simplified form of $T(E)$ for this case is also available [18].

In this paper, we wish to introduce the exponential potential
\begin{equation}
V(x)= V_0( e^{2|x|/a}-1),
\end{equation}
When $V_0>0$, it is an open well which diverges to $\infty$ as $|x|\rightarrow \infty$, see Fig 1(a). When $V_0<0$ it represents a bottomless  potential barrier (Fig. 1(b)). In the following, we  discuss the extraction of bound state eigenfunctions and eigenvalues and extraction of scattering states giving rise to reflection $r(k)$ and transmission $t(k)$ amplitudes from exact analytic solutions of Schr{\"o}dinger equation. We also find them by using the semi-classical approximation  called WKB method. An instructive connection of bound state eigenvalues with the poles of the reflection and transmission coefficients will also be discussed.
\begin{figure}[ht]
	\centering
	\includegraphics[width=7cm,height=5 cm]{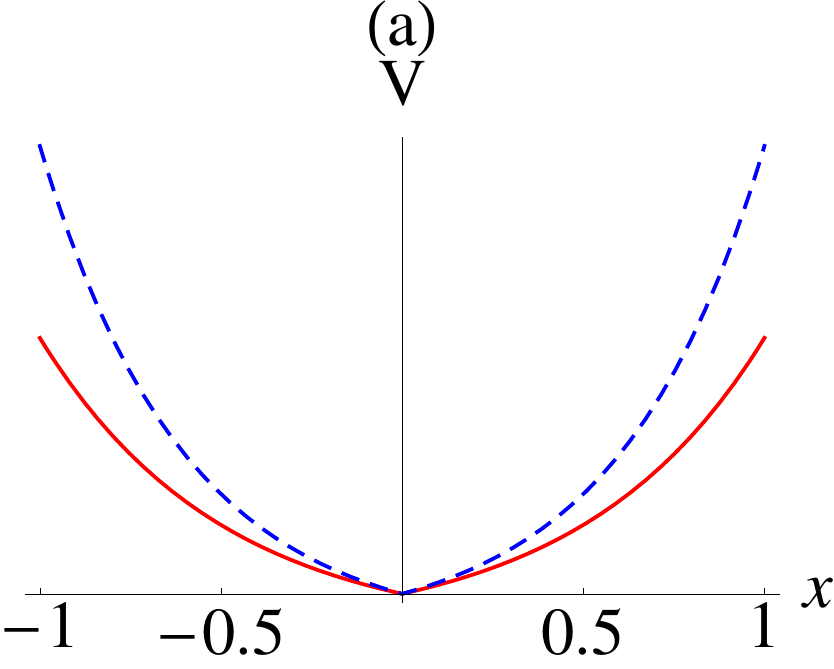}
	\hskip 1 cm
	\includegraphics[width=7cm,height=5 cm]{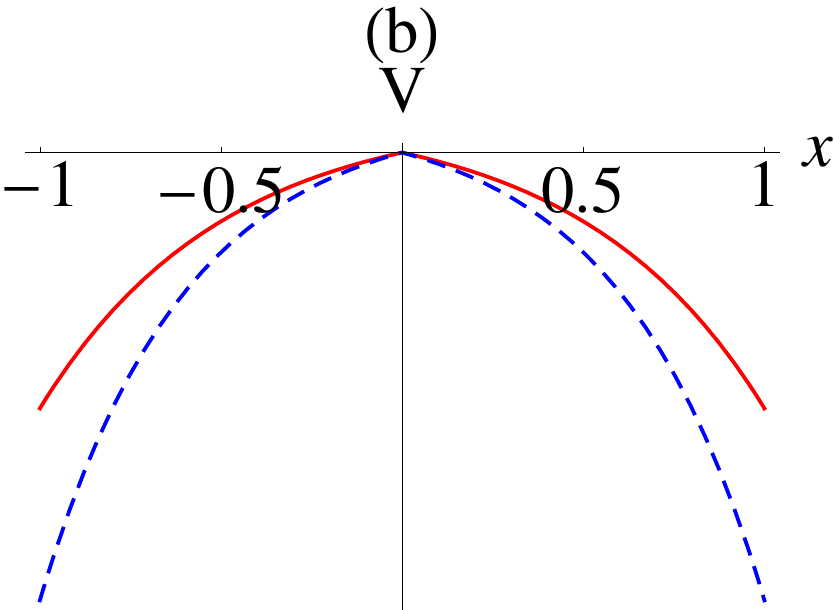}
	\caption{Plots of the exponential potentials: (a) the open well (Eq.(1), $V_0=1$) and (b) the bottomless barrier (Eq.(8), $U0=1)$. For the length parameter $a=1$ see red/solid lines and for  $a=0.8$ see blue/dashed, for larger values of $a$ these potentials are  thicker. The open well has only discrete positive energies as bound state eigenvalues. The barrier has only scattering states at both positive and negative values.}
\end{figure}	

\section{Boundstates in the open well}
We write the Schrodinger equation for (1) as
\begin{equation}
\frac{d^2\psi(x)}{dx^2}+[k^2-q^2e^{2|x|/a}]\psi(x)=0, \quad k=\sqrt{2\mu (E+V_0)}/\hbar, \quad q=\sqrt{2\mu V_0}/\hbar. 
\end{equation} 
By using the transformation $z=\lambda e^{|x|/a}$ [19] in (2) we can transform it to modified Bessel equation [20] as 
\begin{equation}
z^2\frac{d^2\psi(z)}{dz^2}+z\frac{d\psi(z)}{dz}-(\nu^2+ z^2) \psi(z)=0, \quad \nu=ika, \quad \lambda=qa.
\end{equation}
This second order equation is known to have two linearly dependent solutions as $I_\nu(z)$ and $K_\nu(z)$. Despite $\nu=ika$ being imaginary both $I_{\pm ika}(\lambda e^{|x|/a})$ [20] are real continuous functions of $x$. When $x$ and so $z$ is large $I_{\nu}(z)$ diverges $\sim \frac{e^z}{\sqrt{2\pi z}}$ but $K_{\nu}(z)$ converges to zero as $\sim \frac{e^{-z}}{\sqrt{2\pi z}}$. Thus, it is the second solution that satisfies Dirichlet condition that $\psi(\pm \infty)=0$. Since the potential (1) is symmetric, the Eq. (2,3) will have definite parity even/odd solutions. The even parity solution is
\begin{equation}
\psi_e(x)= A K_{ik_n a}(qa e^{|x|/a}), \quad K'_{ik_n a}(qa)=0, \quad  n=0,2,4,...
\end{equation}
The condition that $K'_{ik_n a}(qa)=0$ quantizes energy $E=k^2_n$ and ensures the differentiability of $\psi(x)$ at $x=0$ despite $|x|$ being there. The odd parity solution is given as
\begin{equation}
\psi_o(x)= A~ \mbox{sgn}(x)~ K_{ik_n a}(qae^{|x|/a}),\quad K_{ik_na}(qa)=0, \quad n=1,3,5,...
\end{equation}
The condition that $K_{ik_n a}(qa)=0$ quantizes energy $E=k^2_n$ and ensures that these (odd) eigenfunction vanish essentially at $x=0$. Eqs (4) and (5) give complete spectrum of the open
well which has infinite number of eigenvalues.

For a demonstration of first four bound states of the well, let us take $V_1=1,a=1$ in arbitrary units where  $2\mu=1=\hbar^2$ for convenience. This choice corresponds to a particle whose mass is roughly 4 times that of mass of electron; mass/energy and length are measured  in eV and $A^0$ (Angstrom), respectively. This choice also corresponds to a particle whose mass is roughly 20 times  that of proton or neutron when mass/energy and length are measured in MeV and fm (Fermi), respectively. 

We plot $K_{ik}(1)$ and $K_{ik}'(1)$ ($qa=1$) as a function of $E$ where $k=\sqrt{E+1}$ to find that these functions pass through zero roughly around $E \sim 2,13$ and $E\sim 8,19$, respectively. Next, we find 
exact roots by using ``FindRoot" of ``mathematica" using these four rough guess values. We find the exact eigenvalues are $E_0=2.6759$ and $E_2=13.3305$ for even states with nodes as 0 and 2, respectively. The exact eigenvalues of odd states as $E_1=7.7766$ and $E_3 = 19.5616$ having 1 and 3 nodes, respectively. These four eigenstates are plotted in Fig. 2. Students will find it interesting to check these eigenvalues and eigenstates by using a simple one line  ``mathematica" program called ``wag-the-dog" method  (see Problem 2.54 on page 104 in Ref. [1]) for symmetric potential wells.

\begin{figure}[t]
	\centering
	\includegraphics[width=3.8cm,height=6 cm]{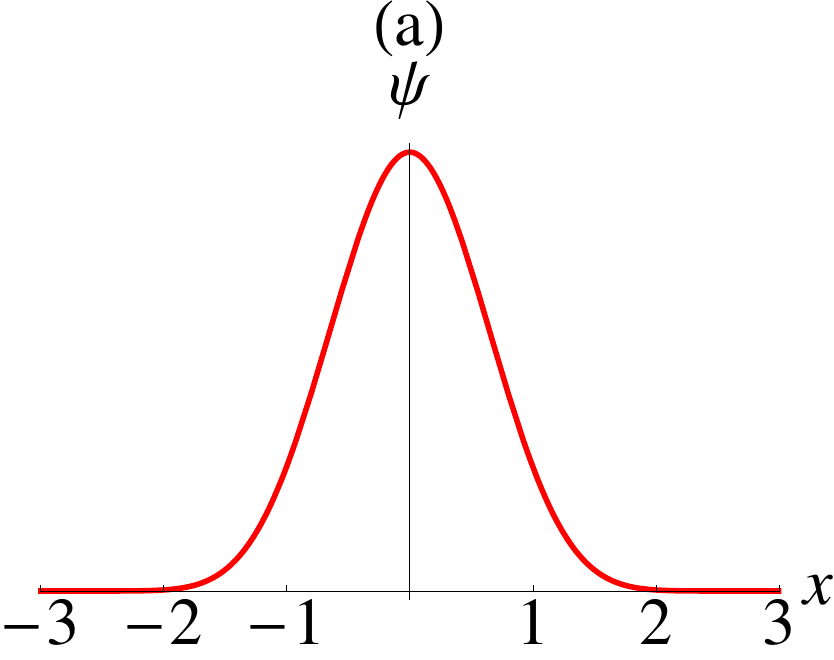}
	\hskip .2 cm
	\includegraphics[width=3.8 cm,height=6 cm]{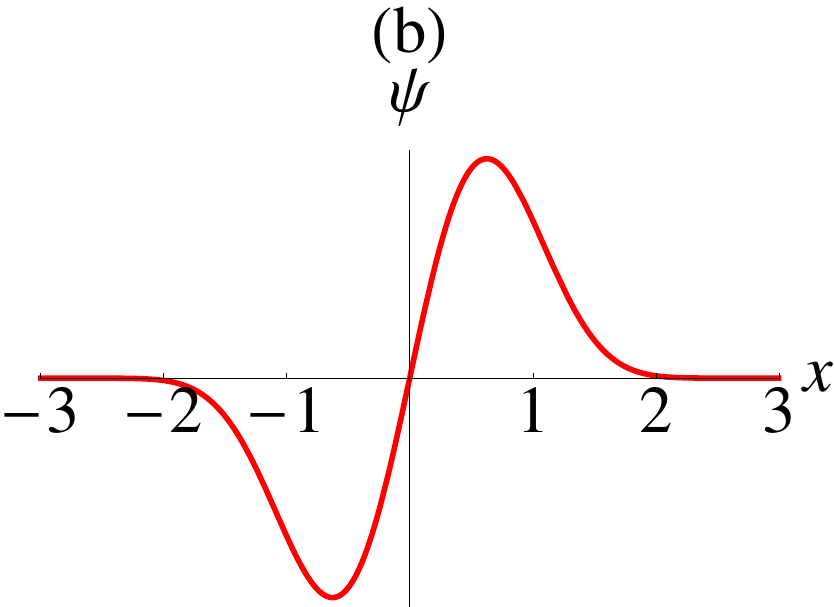}
 	\hskip .2 cm
	\includegraphics[width=3.8 cm,height=6 cm]{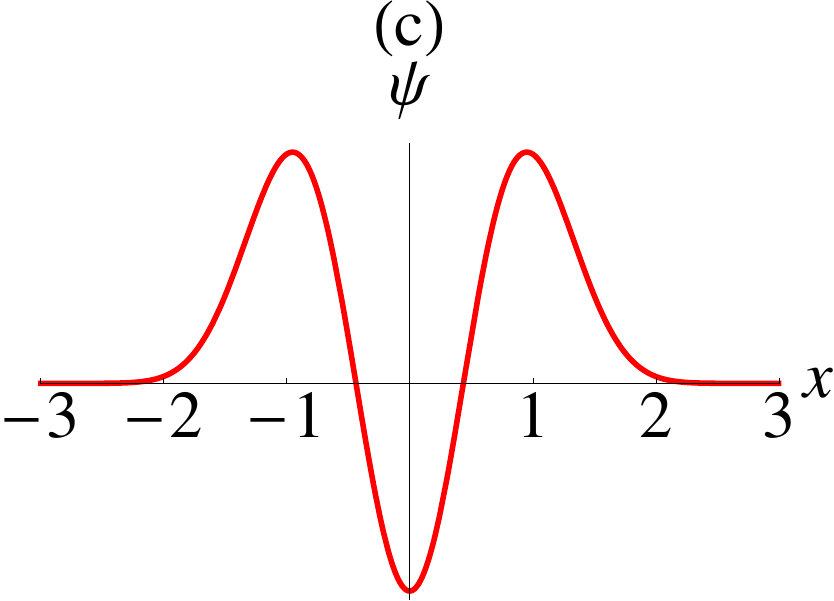}
	\hskip .2 cm
	\includegraphics[width=3.8 cm,height=6 cm]{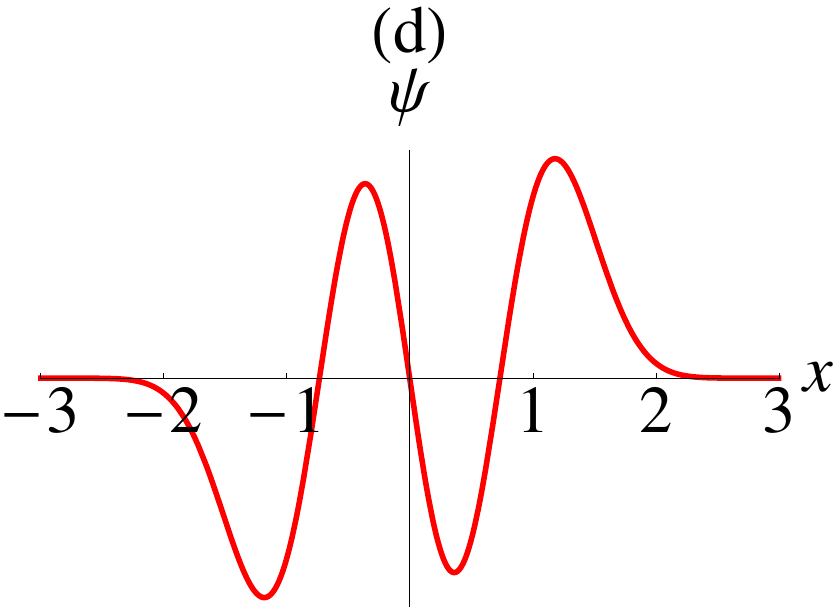}
	\caption{First four (a-d) un-normalized definite  parity bound state eigenfunctions (4,5) for the open well (1) are shown.  The corresponding energy eigenvalues are $E_0=2.6759, E_1=7.7766, E_2=13.3305$ and $E_3=19.5616$}
\end{figure}
\vskip -.5 cm	
\section{Bound state eigenvalues by semi-classical quantization}
A potential well $V(x)$  having a minimum and two real classical turning points $x_1(E),x_2(E)$ at positive/negative energies may have real discrete eigenvalues provided
\begin{equation}
\frac{1}{\pi}\int_{x_1(E)}^{x_2(E)} \sqrt{2\mu[E-V(x)]} dx = (n+1/2)~ \hbar, \quad n=0,1,2,3..., V(x_1)=E=V(x_2).
\end{equation}
Eq. (6) is known as Bohr-Sommerfeld phase space quantization ($n$ in place of $n+1/2$) later
it has been derived from Schr{\"o}dinger equation and it is known as WKB approximation [1-5] for bound state eigenvalues. For the potential well (1) $x_{1,2}=\pm \frac{1}{2} \log [(E+V_0)/V_0]$ and the integral (6) can be performed to give 
\begin{equation}
f(E)=\frac{2qag}{\pi}\left (\tanh^{-1}\frac{\sqrt{g^2-1}}{g}-\frac{\sqrt{g^2-1}}{g}\right)=(n+1/2), \quad g=\sqrt{(E+V_0)/V_0}.
\end{equation}
In Fig. 3, $f(E)$ has been plotted along with horizontal lines at $f=0.5,1.5,2.5,3.5$
\begin{figure}[ht]
	\centering
	\includegraphics[width=14cm,height=7 cm]{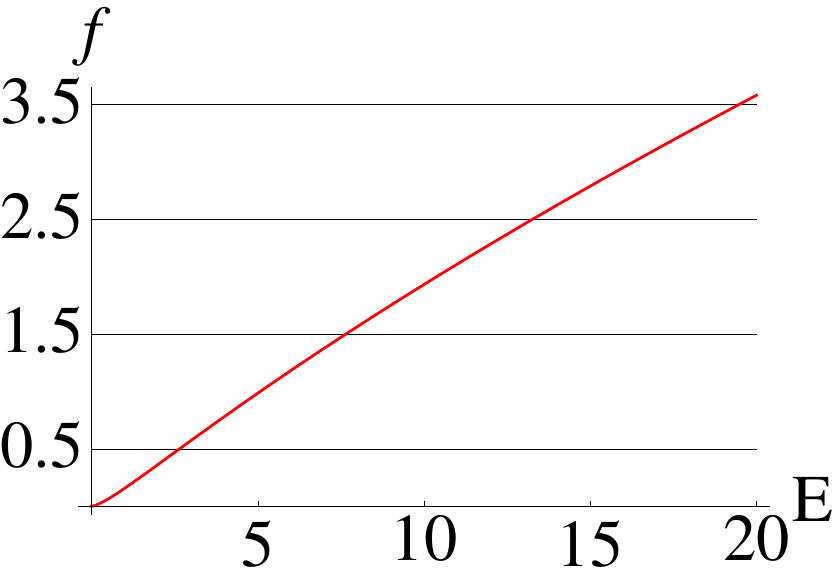}
	\caption{$f(E)$ (7) plotted for $V_0=1$ and $a=1$, $f=0.5,1.5,2.5,3.5$ cut x-axis at 
	$E= 2.6471, 7.6486, 13.2651$ and $19.4743$, these semi-classical eigenvalues lie very close to the exact ones listed in Fig. 2}
\end{figure}	
which cut the energy axis at  $2.6471, 7.6486, 13.2651, 19.4743$. These are  first four 
semi classical eigenvalues which are for the case $V_0=1$ and $a=1$. It can be seen that they lie very close to $E_0,E_1,E_2,E_3$ obtained above using the exact quantum condition (4,5) for bound states. 
\section{Scattering states in bottomless barrier}
The bottomless exponential potential barrier can be written as
\begin{equation}
V(x)=-U_0 (e^{2|x|/a}-1), \quad U_0>0,
\end{equation}
see Fig 1(b). $E=0$ marks  the top of the barrier so this potential has only scattering states 
at both positive and negative energies. The Schr{\"o}dinger equation for this potential is written as
\begin{equation}
\frac{d^2\psi(x)}{dx^2}+[p^2+s^2e^{2|x|/a}]\psi(x)=0, \quad p=\sqrt{2\mu (E-U_0)}/\hbar, \quad s=\sqrt{2\mu U_0}/\hbar. 
\end{equation} 
Again using the transformation $z=\lambda e^{|x|/a}$ [19] in (9) we can transform it to modified Bessel equation as 
\begin{equation}
z^2\frac{d^2\psi(z)}{dz^2}+z\frac{d\psi(z)}{dz}+(-\nu^2+ z^2) \psi(z)=0, \quad \nu=ipa, \quad \lambda=sa.
\end{equation}
 This cylindrical Bessel equation whose two pairs of linearly independent solutions are $J_{\pm \nu}(z)$ and the Hankel functions $H^{(1)}_{\nu} (z)$ and  $H^{(2)}_{\nu} (z)$ [20], any member of these pairs can be expressed as a linear combination of other two. However the latter pair is directly useful since for large values of $x$ or $z$  the Hankel functions [20] behave as scattering states:
 \begin{equation}
H^{(1,2)}_{\nu}(z)\sim \sqrt{2/(\pi z)}e^{\pm i (z-i\nu\pi/2-\pi/4)}.
\end{equation}  
For our potential barrier Fig. 1(b), when the potential is real the scattering is independent of the direction of the incidence of a particle or wave at the potential. So without a loss of generality let us choose the direction of incidence from left. Thus, for scattering from left the incident, reflected and transmitted are to be chosen from
 $\psi_1(x)=H^{(2)}_{ipa}(sa e^{-x/a})$, $\psi_2(x)=H^{(1)}_{ipa}(sa e^{-x/a})$ and $\psi_3(x)=H^{(1)}_{ipa}(sa e^{x/a})$. 
  For a scattering  solution $\psi(x)=\alpha(x) e^{i\beta(x)}$ the time dependent state is  written
 as $\Psi(x,t)=\alpha(x) e^{-i\Phi(x,t)}=\alpha(x)e^{i(\beta(x)-Et/\hbar)}$. Then the  sign of the phase velocity $dx/dt$  determined by the condition of stationary phase $\frac{d\Phi(x,t)}{d t}=0$ at an asymptotic distance decides the relative direction of running of these waves arising from solutions $\Psi_m(x,t) (m=1,2,3)$. So using (11) let us write the asymptotic forms as
\begin{eqnarray}
 \Psi_1(x,t)& \sim & \sqrt{2/(\pi z)} e^{i[-sa e^{-x}-Et/\hbar]} \Rightarrow dx/dt>0, \Psi_2(x,t)  \sim  \sqrt{2/(\pi z)} e^{i[sa e^{-x}-Et/\hbar]} \Rightarrow dx/dt<0, \nonumber\\ \Psi_3(x,t) & \sim & \sqrt{2/(\pi z)} e^{i[sa e^{x}-Et/\hbar]} \Rightarrow dx/dt>0.
\end{eqnarray}
Therefore $\Psi_1$, $\Psi_3$ represent waves moving left to right to be  designated as
incident $\psi_i(x)$ and transmitted $\psi_t(x)$ waves, respectively. The reflected wave needs to be traveling opposite to the incident wave and also it should  be time reversed (complex conjugate) form of the incident wave, so we choose $\psi_2(x)= H^{(1)}_{-ipa}(sa)$ (notice - sign in the subscript) as reflected $\psi_r(x)$ wave.
\begin{figure}[h]
	\centering
	\includegraphics[width=7cm,height=7 cm]{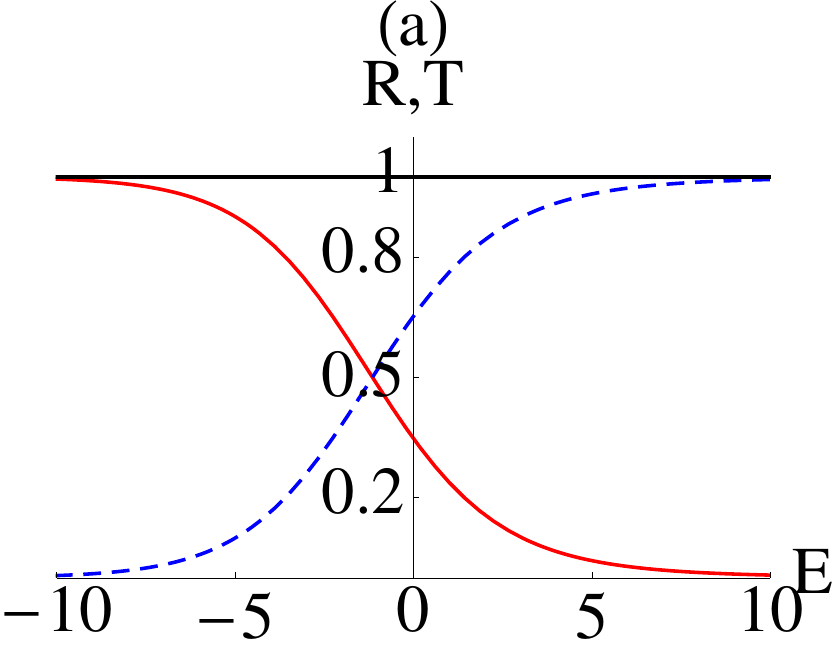}
	\hskip .5 cm
	\includegraphics[width=7cm,height=7 cm]{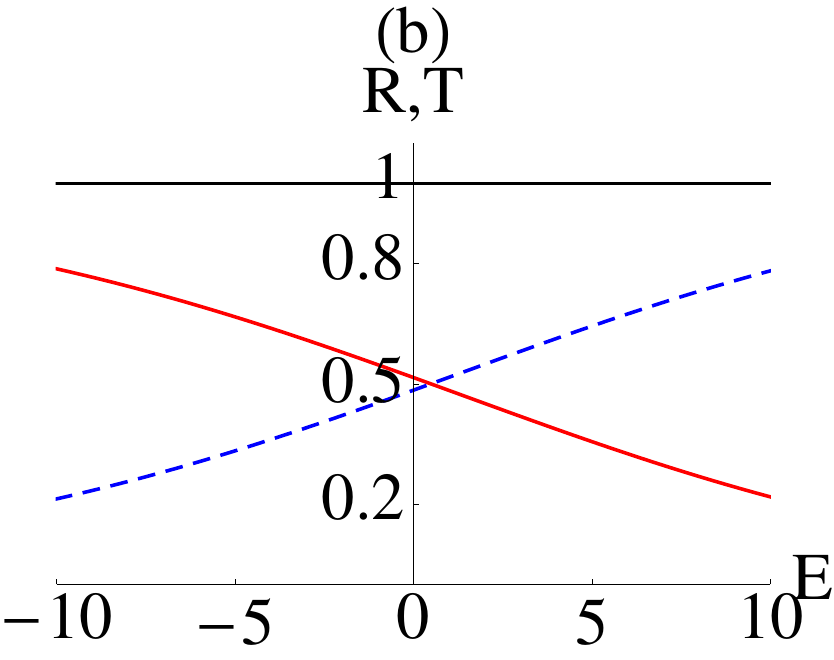}
	\caption{The reflection $R(E)$ (red/solid) and transmission $T(E)$ (blue/dashed) probabilities as a function of 
		energy (15,19) when a particle or wave is incident on the unbounded potential barrier (8) ($U_0=5$ and (a): $a=1$, (b) $a=0.2$). The cross-over $R=T=1/2$ occurs at $E=-1.1487<0$. Normally this result is expected at energy  equal to the top of the barrier namely $E=0$. However for low  values of the effective parameter $qa$ the cross over between $R(E)$ and $T(E)$ can be made to occur at zero or positive energies for instance if $a=0.2$, the crossover occurs at $E=0.4886>0.$ See part (b).}
\end{figure}
Note that $[H^{(2)}_{i\nu}(z)]^*=H^{(1)}_{-i\nu}(z)$ [20]. We can now write the full solution of (9)  as
 \begin{equation}
\psi(x<0) = A H^{(2)}_{ipa}(sa e^{-x/a}) + B H^{(1)}_{-ipa}(sa e^{-x/a}), \quad \psi(x>0)= C H^{(1)}_{ipa}(sa e^{x/a}).
 \end{equation}
The condition of continuity and differentiability of $\psi(x)$ everywhere demands the above two pieces of solutions to be continuous and and differentiable at $x=0$. So we get
\begin{equation}
A H^{(2)}_{ipa}(sa)+ B H^{(1)}_{-ipa} (sa)= C H^{(1)}_{ipa}(sa), \quad A H^{(2)'}_{ipa}(sa) + B H^{(1)'}_{-ipa} (sa)= -C H^{(1)'}_{ipa}(sa)
\end{equation}
We get expressions of A,B and C by solving (14). We get $A=2  H^{(1)}_{-ipa}(sa) H^{(1)'}_{ipa}(sa)$, $B= H^{(1)'}_{ipa}(sa) H^{(2)}_{ipa}(sa)+ H^{(1)}_{ipa}(sa) H^{(2)'}_{ipa}(sa)$, $C=H^{(1)}_{-ipa}(sa) H^{(2)'}_{ipa}(sa)-H^{(1)'}_{-ipa}(sa) H^{(2)}_{ipa}(sa).$  Consequently the ratios $B/A$ and $C/A$ turns out to be
are obtained as 
\begin{eqnarray}
\frac{B}{A}&=&-\frac{1}{2} e^{\pi pa}\left [\frac{H^{(2)}_{ipa}(sa)}{H^{(1)}_{ipa}(sa)}+\frac{H^{(2)'}_{ipa} (sa)}{ H^{(1)'}_{ipa}(sa)} \right]
\nonumber \\
\frac{C}{A}&=&\frac{2i/(\pi sa)}{H^{(1)}_{ipa}(sa)~ H^{(1)'}_{ipa}(sa)}.
\end{eqnarray}
In above, we have used the properties that $H^{(1)}_{-\nu}(z)=e^{i \pi\nu} H^{(1)}_{\nu}(z)$,
$H^{(2)}_{-\nu}(z)=e^{-i\pi \nu} H^{(2)}_{\nu}(z)$ and $[H^{(1)}_{\nu}(z) H^{(2)'}_{\nu}(z)-H^{(1)'}_{\nu}(z) H^{(2)}_{\nu}(z)]=-4i/(\pi z) [20].$ These properties also help
in calculating the current density [1-5]
\begin{figure}[h]
	\centering
	\includegraphics[width=7cm,height=7 cm]{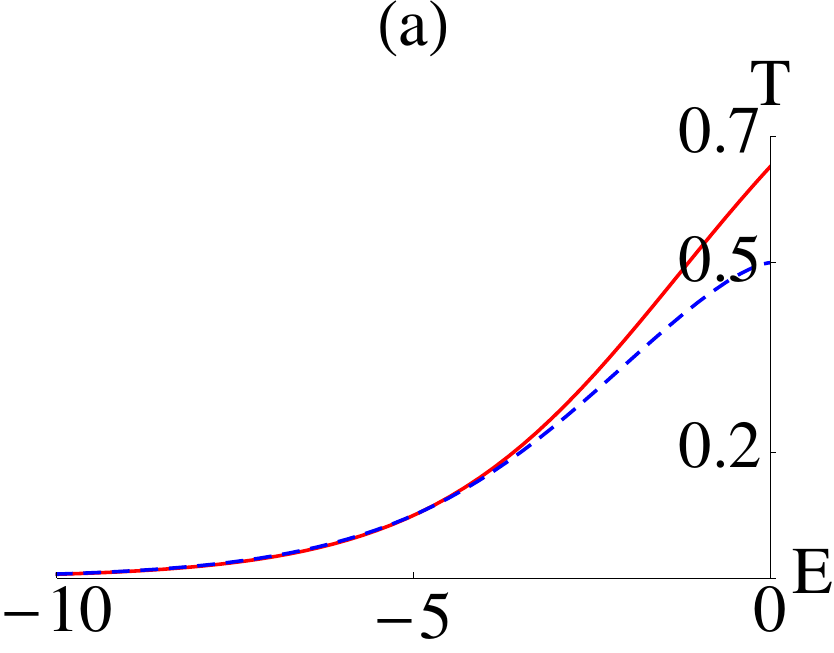}
	\hskip .5 cm
	\includegraphics[width=7cm,height=7 cm]{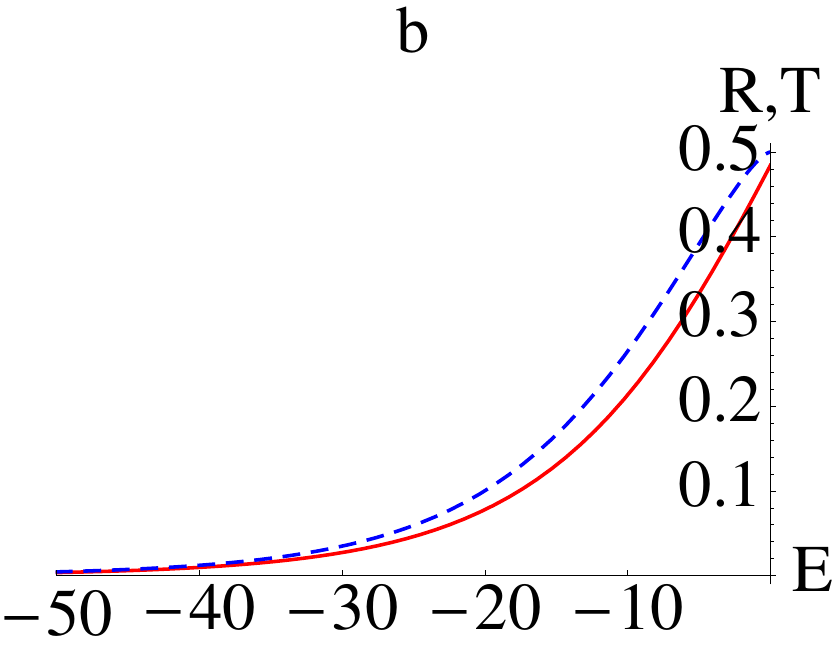}
	\caption{Exact (Eqs. (15,19), solid)  and WKB (Eq.(21), dashed) transmission probabilities for (a): $U_0=5, a=1$, (b): $U_0=5,a=0.2.$ In (a) $T(0)>T_{WKB}(0)=1/2$ and in (b) $T(0)<T_{WKB}(0)=1/2.$ $E=0$, marks the barrier-top }
	\end{figure}
\begin{equation}
{\cal J}=\frac{\hbar}{2i\mu}[ \psi^*(x) \psi'(x)-\psi(x) \psi^{*'}(x)]
\end{equation}
easily on the left and right due to $\psi(x)$ in (13). We get when $p$ is purely imaginary
\begin{equation}
{\cal J}_<={\cal J}_0 (AA^*-BB^*),\quad  \mbox{and}, \quad {\cal J}_>={\cal J}_0 CC^*, \quad {\cal J}_0=\frac{2\hbar}{\mu a \pi},
\end{equation}
on the left and the right of $x=0$. But when $p$ is real, we get
\begin{equation}
{\cal J}_<={\cal J}_0 e^{-\pi p a} (AA^*-BB^*), \quad {\cal J}>={\cal J}_0 e^{\pi p a}CC^*,
\end{equation}
on the left and right of $x=0$. The conservation of particle flux at or continuity of current density $x=0$ is met as $1-R(E)=T(E)$, where $R(E)$
and $T(E)$ reflection and transmission probabilities of the particle incident at the barrier (8).
Thus $R(E)$ and $T(E)$ for the barrier (8) can  be given commonly as
\begin{equation}
R(E)=\left|\frac{B}{A}\right|^2, \quad T(E)=\left|e^{\pi pa} \frac{C}{A}\right|^2,
\end{equation}
where we use the results in eq. (15). Alternatively simpler way is to calculate ${\cal J}_i$, ${\cal J}_r$ and ${\cal J}_t$ may be calculated using $A\psi_i, B \psi_r$ and $C\psi_t$ in (16)
individually as  ${\cal J}_i={\cal J}_0 AA^* , {\cal J}_r=-{\cal J}_0 BB^*$ and ${\cal J}_t={\cal J}_0 CC^*$, when $p$ is imaginary.  But when $p$ is real, the three current densities are ${\cal J}_i={\cal J}_0 e^{-\pi pa}AA^*$, ${\cal J}_r=-{\cal J}_0 e^{-\pi pa} BB^*$ and ${\cal J}_t={\cal J}_0 e^{\pi pa}CC^*$, respectively. Next we define  $R(E)=|{\cal J}_r/{\cal J}_i|$ and  $T(E)=|{\cal J}_t/{\cal J}_i|$ to get (19) again.

We take $U_0=5$ and plot $R(E)$ and $T(E)$ using (15,19) in Fig. 4(a,b), for two cases, (a): $a=1$ and (b): $a=0.2$. The unitarity condition that $R+T=1$ is satisfied at every energy (positive or negative). This is one stringent test of consistency of the obtained results (15,19). Notice that for $a=1$ when the barrier is thicker (see fig. 1(a)) $R(E)$ and $T(E)$ cross over at enrergy $E=E_c=-1.1487<0$ (below the top of the barrier). But in the case of a thinner barrier $(a=0.2$)  the cross-over occurs at an energy $E_C=0.4886>0$. One can also arrange $a>0.2$ so that the cross-over occurs almost at $E=E_c=0$ just at the barrier-top.

\section{WKB approximation for $T(E)$}
The classical turning points for energies $E<0$ for the barrier (8) are given as $x_{1,2}=\pm\frac{1}{2} \log(1-E/U_0)$. Due to non-differentiability of $V(x)$, $x_{1,2}$ for energies $0<E<U_0$ loose meaning as we get $|x|=\frac{1}{2} \log(1-E/U_0)<0$. We can therefore use
WKB approximation for $E<0$ (below the barrier), where we calculate $T(E)$ [1-5] as
\begin{equation}
F=\int_{x_1}^{x_2} \sqrt{\frac{2\mu}{\hbar^2}[V(x)-E]} ~dx, \quad T_{WKB}(E)=\frac{1}{1+\exp(2F)}~,\quad E<0
\end{equation}
For the potential barrier (8) we find
\begin{equation}
F=2saG\left({\tanh^{-1}}\frac{\sqrt{G^2-1}}{G}-\frac{\sqrt{G^2-1}}{G}\right), \quad G=\sqrt{1-\frac{E}{U_0}}, \quad s=\frac{\sqrt{2\mu U_0 a^2}}{\hbar}.
\end{equation}
In Fig. 5, we plot $T(E)$ arising due to exact (Eq. (19), solid)  and WKB (Eq. (21), dashed) approximation for the exponential barrier (8). The WKB seems to work at deep sub-barrier (much beloww the barrier) energies where the barriers are thicker. Near the top of the barrier at $E=0$, the WKB method (21) under-estimates (over-estimates) transmission probability for thicker (thinner) barrier. 
\section{Discussion}
An interesting survey of $R(E)$ and $T(E)$ of solvable potential barriers can be done with regard to the cross over energy $E_c$ where these two probabilities cross each other to have 
$R(E_c)=1/2=T(E_c)$. For the Dirac delta $V(x)=V_0 \delta(x)$ [1], $E_c=V_0/4 <V_0$, For square barrier [1-5] one can have $E_c=V_0$, when $qa=1$ and $E_c$ is both $> (<)$ $V_0$ according as $a>1 (a<1)$. The same experience can be had using the exact $T(E)$ for the Eckart $V(x)= V_0 \mbox {sech}^2(x/a)$ [3,4], exponential $V(x)=V_0 e^{-|x|/a}$ [9] and the  Morse [11] barriers which are bounded barriers. For the  bottomless  parabolic barrier $V(x)=V_0(1-x^2/a^2)$ [4,5], $E_c=V_0$ strictly and irrespective of the values of $a$. However, for the exactly solvable triangular barrier $V(x)=V_0(1-|x|/a)$ [5], $T(E=V_0)=3/4$ so $E_c<V_0$, irrespective of the values of $a$. We find that for the bottomless exponential potential, like the cases bounded barriers, all three cases could be arranged by varying the length parameter $a$. See Fig. 4(a) and 4(b) for $a=1$, $E_c$ is negative (below the barrier) and for $a=0.2$, $E_c>0$ (above the barrier). Notice that the cross-over of $T(E)$ and $R(E)$ for three bottomless barriers presents three different scenarios. This underlies the importance of studying exactly solvable cases which bring out such disparate results, more so when they diverge as $|x|\rightarrow \infty.$ Interestingly, the WKB approximation gives $R=1/2=T$ at the barrier top energy (incorrectly) in all three cases of bottomless barriers mentioned here.

\begin{figure}[t]
	\centering
	\includegraphics[width=14cm,height=7 cm]{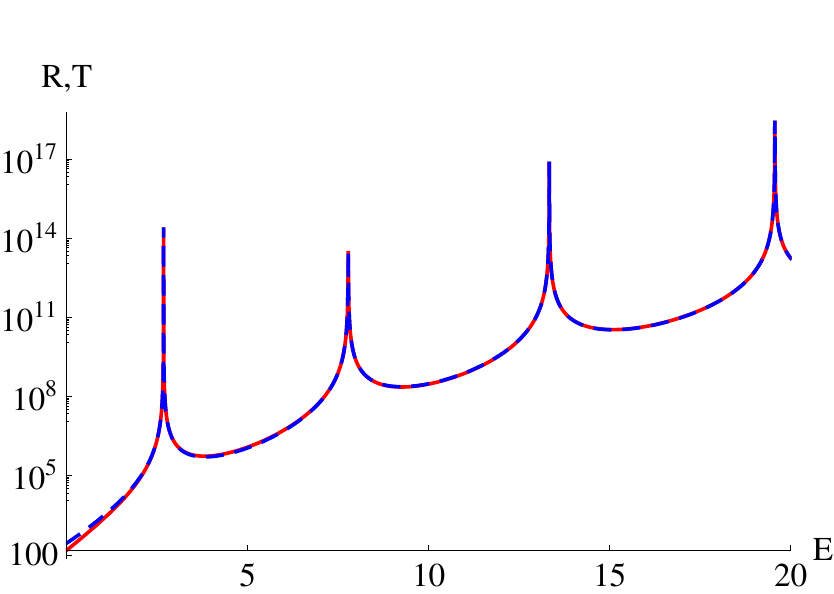}
	\caption{The poles in $R(E)$ (red/solid) and $T(E)$ (blue/dashed) at the bound state eigenvalues (see Fig. 3) of the open well, when we change $U_0 \rightarrow -V_0$ in Eqs. (15,19). Here again we take $V_0=1$ and $a=1$.}
\end{figure}
Poles of reflection $r(k)=B/A$ and transmission $t(k)=C/A$ amplitudes, of the type $k=i\alpha$ ($\alpha=\sqrt{-E}$) are known to yield the discrete energy bound states of a potential well which converges to zero asymptotically (e.g. square well, Gaussian well). One may extend this idea to  extract the bound state eigenvalues from the poles of $R(E)$ and $T(E)$. In a textbook [2], this has been done for the square well potential but unfortunately the negative energy poles of these coefficient have been shown as having $T=1$,  whereas they should have been shown as much higher thin spikes representing $T=\infty$. 

In Eq.(15),  let us replace  $U_0$ by $-V_0$,  then $s\rightarrow iq$ and Hankel function  $H^{(1)}_{ipa}(sa)$ converts to the modified Bessel function $K_{ipa}(qa)$ due to the property [20] that $H^{(1)}_{\nu}(iz)=-{2i}{\pi} e^{-\nu i\pi/2}K_{\nu}(z)$. In this way, the common denominator of $r(E)=B/A$ and $t(E)=C/A$ (15) yields
the poles as 
\begin{equation}
K_{ipa}(qa)~ K'_{ipa}(qa)=0, \quad  p= \sqrt{2\mu(E+V_0)}/\hbar
\end{equation}
which is the combined bound state eigenvalue condition derived above (4,5) for the open potential well (1). Thus, if we plot $R(E)$ and $T(E)$ (15,19) by changing $U_0$ to $-V_0$ as in 
Fig. 5 for $V_0=1, a=1$, we get four poles in them at around $E \sim 2,7, 13$ and 19, which are capable of yielding the correct bound state eigenvalues of the open well as mentioned in Fig. 2. It may be remarked  that the thus changed expressions of $R(E)$ and $T(E)$ may not be of any use as the tunneling through open potential wells does not make any sense, however, these changed coefficients can yield the possible bound states of the well. Further,  one may connect
Eqs. (7) and (21) in this regard. When we change $U_0\rightarrow -V_0$ in (21), $s \rightarrow iq$, $G \rightarrow g$, $F\rightarrow f$ and $T_{WKB}(E)\rightarrow [1+e^{2if}]^{-1}$, one can readily get poles of $T_{WKB}(E)$ as $F=(n+1/2)\pi$ which is nothing but Eq. (7). We feel that such an interesting connection is often not discussed in textbooks.

\section{Conclusion}
Lastly, we hope that the new open well and the bottomless barrier constructed from the exponential potential will be welcome as solvable models which are rich in displaying several interesting quantum mechanical features through  commonly available properties of Bessel and Hankel functions. Among the three (parabolic, triangular and exponential) analytically solvable bottomless barriers, the exponential barrier presented here has the new feature that the cross-over of the reflection and transmission probabilities can occur at the barrier-top and also either below a thicker barrier or above a thinner barrier. This is generally true for bounded potential barriers which vanish asymptotically on one or both sides. The simple WKB approximation does not yield this interesting result and the numerical integration of Schr{\"o}dinger equation is not plausible due to the lack of generic asymptotic boundary conditions, it requires analytic solutions in every particular case.

\end{document}